\documentclass[12pt,a4paper,final]{iopart}
\pdfoutput=1
\usepackage{graphicx}
\usepackage[breaklinks=true,colorlinks=true,linkcolor=blue,urlcolor=blue,citecolor=blue]{hyperref}
\usepackage{amsfonts, amsmath, amssymb, dsfont}

\usepackage{leftidx}
\usepackage{cite}
\usepackage{booktabs}
\usepackage{wasysym}
\usepackage{esint}

\usepackage[polish,danish,english]{babel}

\def\limth{\lim\nolimits_\text{th}}

\def\boldx{{\boldsymbol{x}}}
\def\blam{{\boldsymbol{\lambda}}}
\def\bmu{{\boldsymbol{\mu}}}

\def\be{\begin{equation}}
\def\ee{\end{equation}}
\def\barr{\begin{IEEEeqnarray}}
\def\earr{\end{IEEEeqnarray}}

\numberwithin{equation}{section}

\newcommand{\nn}{\nonumber}

\def\vtheta{\vartheta}

\begin{document}


\title{Density form factors of the 1D Bose gas for finite entropy states}

\author{J. De Nardis$^1$ and M. Panfil$^2$}

\address{$^1$Institute for Theoretical Physics, University of Amsterdam, Science Park 904,\\
Postbus 94485, 1090 GL Amsterdam, The Netherlands}

\address{$^2$ International School for Advanced Studies (SISSA) and INFN,\\
via Bonomea 265, 34136 Trieste, Italy}

\eads{\mailto{jdenardis@uva.nl}, \mailto{mpanfil@sissa.it}}


\begin{abstract}
We consider the Lieb-Liniger model for a gas of bosonic $\delta-$interacting particles. Using Algebraic Bethe Ansatz results we compute the thermodynamic limit of the form factors of the density operator between finite entropy eigenstates such as finite temperature states or generic non-equilibrium highly excited states. These form factors are crucial building blocks to obtain the thermodynamic exact dynamic correlation functions of such physically relevant states. As a proof of principle we compute an approximated dynamic structure factor by including only the simplest types of particle-hole excitations and show the agreement with known results.
\end{abstract}

\maketitle

\section{Introduction}

Integrable quantum models are gaining an increasing role in modern physics. The access to an exact solution of a many-body interacting system gives an unprecedented opportunity to explore strongly correlated quantum systems beyond perturbative or numerical methods. Circumstances are especially encouraging in one-dimensional systems where some integrable models naturally appear and the theoretical predictions are of huge experimental value. For example the recent progresses in cold atoms experiments in optical lattices allow the manufacturing of different models in an almost ideally isolated environment. Many physical properties of some paradigmatic integrable models as the XXZ spin chain and the Lieb-Liniger model were indeed observed in such experiments\cite{2004_Paredes_NATURE_429,2008_Amerongen_PRL_100,2014_Fabbri_arXiv} and similar observations were also possible in highly spatially anisotropic crystals \cite{2013_Mourigal_NatPhys_9,2013_Lake_PRL_111}.

The standard technique to solve interacting integrable models is the Bethe Ansatz\cite{KorepinBOOK}. This technique provides us a complete characterization of eigenstates and, at least in some cases, the matrix elements of physical operators between two eigenstates ({\em form factors}). However these expressions are usually cumbersome since they depend on all the $N$ variables (\emph{rapidities}) describing the eigenstates, where $N$ is the number of constituents in the system. Moreover in order to address the two-point correlation functions one has to perform a summation of form factors over the whole Hilbert space. Performing such summations is so far beyond our analytical abilities but they can be evaluated numerically \cite{2005_Caux_PRL_95,2006_Caux_PRA_74,PhysRevA.89.033605}. In the conformal limit the summations simplifies allowing for an exact treatment \cite{1742-5468-2012-09-P09001,1742-5468-2011-12-P12010}. This led to a derivation of the Luttinger liquid \cite{1981_Haldane_PRL_47,GiamarchiBOOK} (an effective theory of gapless 1D models) results for correlation functions directly from a microscopic (integrable) theory. Similar results were also worked out in this direction: the phenomenological quantities entering the Luttinger liquid description of correlation functions  were connected with microscopic data \cite{2011_Shashi_PRB_84,2012_Shashi_PRB_85}. However the full determination of correlation functions resisted so far the best efforts.

One of the main difficulty arises from a complicated structure of form factors as they are highly non-trivial functions of rapidities of two eigenstates. The thermodynamic limit at fixed density however allows for some simplifications. On the physical grounds we can reason that form factors of local operators are non-zero only when evaluated between two very similar eigenstates. Indeed a local operator is not expected to modify a macroscopic number of degrees of freedom and consequently its form factors are functions of parameters specifying one of the states as an \emph{excited state} over the other which we call an \emph{averaging state}. The number of such excitations is a sub-extensive number $n$  such that in the thermodynamic limit $n/N \to 0$. We can then distinguish two different situations depending on a type of the averaging state. When the averaging state is the ground state of a gapless theory each excitation with a finite momentum and energy gets "dressed" by an infinite number of zero-energy excitations. This sign of the criticality in the system can be seen in a non-integer scaling behavior of the form factors with the size of the system (here the length $L$) as $L^{- \alpha}$, with $\alpha$ a rational number \cite{1990_Slavnov_TMP_82}. This makes the evaluation of the spectral sum, required for computation of dynamical correlation functions, a daunting task. Still some progress was achieved in the aforementioned conformal limit \cite{1742-5468-2012-09-P09001,1742-5468-2011-12-P12010}. Here we focus on another regime, when the averaging state is a finite entropy state. Similar averaging states were already considered, for example, in \cite{1742-5468-2011-01-P01011}.

The excited states of a finite entropy averaging state contribute indeed \emph{individually} to the whole sum over the Hilbert space. This can be seen again at the level of the form factors as their scaling with the system size is determined by number of excitations $n$ as $L^{-n}$. In particular as we restrict here to the (repulsive) Lieb-Liniger model and to the density operator form factors the only relevant excitations are \emph{particle-hole} excitations.

Our main result is the thermodynamic limit of the form factors of the density operator between the finite entropy averaging state and its excited states with a number $n$ of particle-hole excitations. Such form factors are the building blocks of the correlation functions for a general, non-critical situation. The results are applicable to compute correlation functions at finite temperatures (when the averaging state is the thermal state) and also for systems out of equilibrium with the averaging state being the steady state of the unitary time evolution \cite{2013_Caux_PRL_110,PhysRevA.89.033601}.

\subsection{Structure of the article}

In section \ref{1DBoseGas} we recall the Bethe Ansatz solution of the Lieb-Liniger model and we collect all the necessary ingredients to compute the thermodynamic limit of the density form factors. The bulk of the computation is shown in section \ref{thermodynamiclimit}. In section \ref{regularization} we show how to regularize divergences present in form factors. In section \ref{expansion} we compute the dynamical structure factor (density-density correlation function) in the $1/c$ expansion. In section~\ref{numerics} we numerically evaluate the dynamical structure factor by including only the simplest (1 particle-hole) excitations.

\section{1D Bose gas} \label{1DBoseGas}

The Hamiltonian of $N$ bosonic particles confined in a one spatial dimension is  \cite{1963_Lieb_PR_130_1,1998_Olshanii_PRL_81}
\begin{align}\label{H}
  H = -\sum_{j=1}^N \partial_{x_j}^2 + 2c\sum_{j<k}^N \delta\left(x_j - x_k\right) ,
\end{align}
where $c$ is the strength of the two-body, repulsive ($c>0$) interactions and we set $\hbar=1$ and $2m=1$. The wavefunctions are superpositions of plane waves \cite{1963_Lieb_PR_130_1},
\begin{align} \label{wv_fnc}
  \langle \boldx|\blam\rangle = \Psi(\boldx|\blam) = \prod_{j>k}^N \textrm{sgn}\left(x_j-x_k\right)\sum_{P_N} \mathcal{A}_P e^{i\sum_{j=1}^N \lambda_{P_j} x_j},
\end{align}
where the summation extends over all permutations of $N$ particles. We also adopt a shorthand notation in which $\boldx = \{x_j\}_{j=1}^N$ and $\blam = \{\lambda_j\}_{j=1}^N$. The effect of interaction is encapsulated in the coefficients
\begin{align}
  \mathcal{A}_P = (-1)^{[P]} e^{i/2\sum_{j>k} \textrm{sgn}(x_j-x_k)\theta(\lambda_{P_j} - \lambda_{P_k})},
\end{align}
with two-particle phase shift
\begin{align}\label{phase_shift}
  \theta(\lambda) = 2\arctan\left(\lambda/c\right).
\end{align}
The energy of an eigenstate $|\blam\rangle$ is
\begin{align}  \label{energy_discrete}
  E(\blam) = \sum_{j=1}^N \lambda_j^2.
\end{align}
The operator of the total momentum, $\hat{P} = -i\sum_{j=1}^N \partial_{x_j}$, commutes with the Hamiltonian \eqref{H} and its eigenvalues are simply
\begin{align} \label{momentum_discrete}
  P(\blam) = \sum_{j=1}^N \lambda_j.
\end{align}
Imposing the periodic boundary conditions constrains the set of rapidities $\blam$ to solutions of the Bethe equations
\begin{align}  \label{bethe}
  \lambda_j = \frac{2\pi}{L}I_j - \frac{1}{L} \sum_{k=1}^N \theta \left(\lambda_j - \lambda_k\right),
\end{align}
where $L$ is the length of the system. Quantum numbers $I_j$ are integers (half-odd integers) for $N$ odd (even) and follow the Pauli principle - wave function vanishes identically if any two of them coincide. The Hilbert space is spanned by allowed choices of quantum numbers. It is customary to name the eigenstates of \eqref{H} in the finite system the Bethe states. We follow this tradition.

The norm of the Bethe states admits a neat representation in the form of a determinant \cite{1971_Gaudin_JMP_12_I,1982_Korepin_CMP_86}
\begin{align} \label{norm}
  |\blam|^2 \equiv \langle\blam|\blam\rangle = c^N \prod_{j\neq k}^N \frac{\lambda_j - \lambda_k + ic}{\lambda_j - \lambda_k} \det_N \mathcal{G},
\end{align}
where the Gaudin matrix is
\begin{align} \label{gaudin}
  \mathcal{G}_{jk} &= \delta_{jk}\left(L + \sum_{m=1}^N K\left(\lambda_j - \lambda_m\right) \right) - K\left(\lambda_j - \lambda_k\right),\\
  K(\lambda) &= \frac{2c}{\lambda^2 + c^2}.\label{K}
\end{align}
Note also that the kernel $K(\lambda)$ is a derivative of the two-particle phase shift $\theta(\lambda)$.

We define the density operator as
\begin{equation}
\hat{\rho}(x) = \sum_{j=1}^N \delta( x  - x_j)
\end{equation}
where $\{ x_j\}_{j=1}^N$ are the positions of all the particles in the gas. The two-point correlation function of this operator is particularly relevant for both theory and experiment.
The form factors of the density operator are given by the Algebraic Bethe Ansatz approach \cite{1990_Slavnov_TMP_82}
\begin{align} \label{ff}
\langle \bmu | \hat{\rho}(0) | \blam\rangle = \left( \sum_{j=1}^N (\mu_j - \lambda_j) \right)\prod_{j=1}^N \left(V_j^+ - V_j^-\right) \prod_{j,k}^N \left(\frac{\lambda_j - \lambda_k + ic}{\mu_j - \lambda_k}\right)\frac{\det_N \left(\delta_{jk} + U_{jk}\right)}{V_p^+ - V_p^-},
\end{align}
where both $|\blam\rangle$ and $|\bmu\rangle$ are Bethe states. Different factors appearing in \eqref{ff} are
\begin{align} \label{matrix_U}
  V_j^{\pm} &= \prod_{k=1}^N \frac{\mu_k - \lambda_j \pm ic}{\lambda_k - \lambda_j \pm ic},\\
  U_{jk} &= i\frac{\mu_j - \lambda_j}{V_j^+ - V_j^-} \prod_{m\neq j}^N \left(\frac{\mu_m - \lambda_j}{\lambda_m - \lambda_j}\right) \biggl(K\left(\lambda_j - \lambda_k\right) - K\left(\lambda_p - \lambda_k\right) \biggl),
\end{align}
and $\lambda_p$ is an arbitrary number, not necessarily from the set $\blam$.

\subsection{Thermodynamic limit}

We consider now the thermodynamic limit $N\rightarrow\infty$ with fixed density $D = N/L $. We denote such limit with $\lim_{\text{th}}$. The Bethe states can be then characterized by a filling function $\vartheta(\lambda)$ defined as a number of rapidities in an interval $(\lambda,\lambda+d\lambda)$ divided by a maximal number of rapidities (in this interval). Due to an interacting nature of the gas the maximal number of the particles is not constant and the density of particles is connected with the filling function through an integral equation \cite{1969_Yang_JMP_10}
\begin{align} \label{rho_p}
  2\pi \rho(\lambda) &= \vartheta(\lambda)\left(1 + \int_{-\infty}^{\infty} d\mu K(\lambda -\mu) \rho(\mu)\right).
\end{align}
The filling function obeys
\begin{align}\label{filling_condition}
  0 \leq \vartheta(\lambda) \leq 1,
\end{align}
which guarantees the existence of $\rho(\lambda)$ through \eqref{rho_p}.
The filling function provides a complete macroscopic characterization of the Bethe states in the thermodynamic limit. For example the extensive part of the momentum and the energy is (c.f. with eqs. \eqref{energy_discrete} and \eqref{momentum_discrete})
\begin{align}\label{energy_cont}
  P[\vartheta] &= L\int_{-\infty}^{\infty} d\lambda\, \rho(\lambda)\, \lambda.\\
  E[\vartheta] &= L\int_{-\infty}^{\infty} d\lambda\, \rho(\lambda)\, \lambda^2\,.
\end{align}
In this work we focus on the regions of the Hilbert space that are characterized by a smooth (differentiable) filling function and are of the finite energy density: $E[\vartheta]/L < \infty$.

A given smooth filling function corresponds to many different microscopic eigenstates. The number of them is equal to the logarithm of the entropy $S[\vartheta]$, the later is given by \cite{1969_Yang_JMP_10}
\begin{align}
  S[\vartheta] &= L\int_{-\infty}^{\infty} {\rm d}\lambda\ s[\vartheta; \lambda],\\
  s[\vartheta;\lambda] &= \rho_t(\lambda)\log \rho_t(\lambda) - \rho(\lambda)\log \rho(\lambda) - \rho_h(\lambda)\log \rho_h(\lambda),
\end{align}
where we introduced a shorthand notation
\begin{align}
 \rho_t(\lambda) &\equiv \rho(\lambda)/\vartheta(\lambda),\\
 \rho_h(\lambda) &\equiv \rho_t(\lambda)\left(1-\vartheta(\lambda)\right),
\end{align}
where $\rho_t(\lambda)$ has a meaning of the maximal density of the rapidities and $\rho_h(\lambda)$ denotes the density of holes.

The density operator is diagonal in the functional space of the filling functions. Its form factors are nonzero only when the two eigenstates are characterized by the same filling function $\vartheta(\lambda)$ and differ only by a number $n$ of excitations such that $\lim_{\text{th}} \frac{n}{N} = 0$. The density form factors are zero for states containing different number of particles and therefore these excitations occur only as particle-hole pairs. We choose a set $\{ \lambda_j^-\}_{j=1}^n$ of rapidities in the averaging state and we change their values to a new set $\{ \mu_j^+\}_{j=1}^n$. We denote the pair of particle-hole as $\{\mu_j^+, \mu_j^-\}_{j=1}^n$, where $\{ \mu_j^+ \}_{j=1}^n$ are particles: they are not present in the averaging state, while $\{\mu_j^-\}_{j=1}^n$ are holes: they are related to the rapidities we have changed in the averaging state $\{ \lambda^-\}_{j=1}^n$ by $1/L$ corrections as $\lambda_j^- = \mu_j^- + \left(\frac{F(\mu^-)}{L} \right)$ for each $j=1, \ldots , n$ and with the function $F(\lambda)$  defined below. The rapidities $\{ \mu_j^- \}_{j=1}^n$ are absent in the excited state. A single excitation will be denoted here with $\mu^- \to \mu^+$.

Due to the correlated nature of the gas, particle-hole excitations modify the density of particles not only in the vicinity of $\mu_j^+$ and $\mu_j^-$. In fact the density $\rho(\lambda)$ acquires a change of order of $1/L$, as can be seen from studying the difference $\mu_j - \lambda_j$. This difference can be conveniently expressed as \cite{KorepinBOOK}
\begin{align} \label{back flow_def}
  F(\lambda_j) = -L\rho_t(\lambda_j)\left(\mu_j - \lambda_j\right) ,
\end{align}
where $F(\lambda)$ is the back-flow function that fulfills the following linear integral equation (for a single particle-hole excitation)
\begin{align} \label{backflow}
  2\pi F\left(\lambda\,|\, \mu^+, \mu^-\right) =&\; \theta(\lambda-\mu^+) - \theta(\lambda-\mu^-) \nonumber\\
&+ \int_{-\infty}^{\infty} d\mu K(\lambda-\mu) \vartheta(\mu) F\left(\mu\,|\, \mu^+, \mu^-\right) .
\end{align}
The linearity of the back-flow implies that for multiple particle-hole excitations the total back-flow is the sum of individual contributions
\begin{equation} \label{back-sum-flow}
  F\left(\lambda\,|\, \{(\mu_j^+, \mu_j^-)\}_{j=1}^n\right) = \sum_{j=1}^n F\left(\lambda\,|\, \mu_j^+, \mu_j^-\right) .
\end{equation}
It also implies the back-flow can be further factorized in the particle and hole contributions. We define the back-flow for a single excitation as
\begin{align} \label{back-single-flow}
  2\pi F\left(\lambda\,|\, \mu \right) &= \theta(\lambda-\mu)
+ \int_{-\infty}^{\infty} d\gamma K(\lambda-\gamma) \vartheta(\gamma) F\left(\gamma|\, \mu \right) .
\end{align}
This allows to write the momentum and the energy of a single excitation as
\begin{align}
 & k[\vartheta; \mu] = \mu  - \int_{-\infty}^{\infty} \rmd\lambda \vartheta(\lambda) F(\lambda|\mu)\label{exc_momentum} ,\\
  & \omega[\vartheta; \mu] = \mu^2 - 2\int_{-\infty}^{\infty} \rmd \lambda \vartheta(\lambda) \lambda F(\lambda|\mu)\label{exc_energy} ,
\end{align}
which are the fundamental building blocks for the energy and momentum of a thermodynamic state with $n$ particle-holes
\begin{align}
& \Delta \omega = \sum_{j=1}^n \omega[\vartheta; \mu^+_j] - \omega[\vartheta; \mu^-_j] , \nn \\&
 \Delta k = \sum_{j=1}^n k[\vartheta; \mu^+_j] - k[\vartheta; \mu^-_j] .
\end{align}
The excited states, due to the back-flow of their rapidities, have different entropy respect to the averaging state. The difference equals \cite{1990_Korepin_NPB_340}
\begin{align}
 \delta S[\vartheta, \mu^+, \mu^-] = \int_{-\infty}^{\infty} {\rm d}\lambda\, s[\vartheta;\lambda]\frac{\partial}{\partial \lambda}\left(\frac{F(\lambda| \mu^+, \mu^-)}{\rho_t(\lambda)}\right)  \equiv \delta S[\vartheta, \mu^+] -  \delta S[\vartheta, \mu^-] ,\label{S_diff}
\end{align}
where in the last step we used the back-flow function for a single excitation \eqref{back-single-flow}. The differential entropy \eqref{S_diff} corresponds to the number of microstates that share $n$ particle-hole excitations with the same thermodynamic energy and momentum (given by \eqref{exc_energy}  and \eqref{exc_momentum}) but with different sub-leading corrections to them.

Finally, let us introduce the form factors and relate them to the correlation functions. We consider an ensemble average, denoted by $<\cdot>$ of the density-density correlation function. We assume that the ensemble has a saddle-point configuration uniquely specifying the filling function $\vartheta(\lambda)$ \cite{1969_Yang_JMP_10}. In order to compute the form factors, it is useful to directly refer to a specific microscopic configuration that has $\vartheta(\lambda)$ as its thermodynamic limit. We will choose one such finite size configuration and call it a averaging state $|\blam\rangle$. Any other state, with the same filling function but with microscopic differences can be viewed as an excitation (with a positive or negative energy) over the averaging state. The choice of the averaging state is not unique, indeed there is a number $e^{S[\vartheta]}$ of possible choices, but the correlation functions in the thermodynamic limit are independent of this choice for most of the relevant operators \cite{KorepinBOOK}. For simplicity we choose here the averaging state with rapidities distributed such that for each interval $[\lambda, \lambda + d\lambda]$ there are $\rho(\lambda) d\lambda$ uniformly distributed rapidities: $ (\lambda_j- \lambda_k)= \left(\frac{j-k}{L \rho(\lambda_j) } \right)+ \mathcal{O}(L^{-2})$ when $\lambda_j \sim \lambda_k$.

We define then the density density correlation function in the thermodynamic limit~as
\begin{align}
  \langle \hat{\rho}(x,t) \hat{\rho}(0,0)\rangle = \frac{\langle \vartheta|\hat{\rho}(x,t) \hat{\rho}(0,0)|\vartheta\rangle }{\langle \vartheta|\vartheta\rangle} = \lim_{\text{th}} \frac{\langle \blam|\rho(x,t) \rho(0,0)|\blam\rangle }{\langle \blam|\blam\rangle} .
\end{align}
The correlation function in the finite system can be expanded using the complete basis of Bethe states
\begin{align}  \label{corr_func_finite_N}
\frac{\langle \blam|\hat{\rho}(x,t) \hat{\rho}(0,0)|\blam\rangle}{\langle \blam|\blam\rangle} = \sum_{\{\mu_j\}_{j=1}^N} e^{ix(P_{\mu} - P_{\lambda}) -it(E_{\mu} -E_{\lambda})} |\mathcal{F}_N\left( \bmu, \blam\right)|^2.
\end{align}
where we defined the microscopic form factors as
\begin{align} \label{micro_ff}
  \mathcal{F}_N\left( \bmu, \blam\right) = \frac{\langle\bmu| \hat{\rho}(0,0)|\blam\rangle}{\sqrt{\langle \blam|\blam\rangle \langle \bmu|\bmu\rangle}} ,
\end{align}
and we used that
\begin{align}
   \hat{\rho}(x,t) = e^{i \left(Ht - Px\right)} \hat{\rho}(0,0)e^{-i \left(Ht - Px\right)} .
\end{align}
To proceed further in taking the thermodynamic limit it is important to note two things. First the summation in eq.~\eqref{corr_func_finite_N}  is constrained since the set of rapidities must be a solution to the Bethe equations \eqref{bethe}. On the other hand the form factors become in the thermodynamic limit rather smooth functions of the filling $\vartheta(\lambda)$ and of the particles-holes momenta $\{\mu_j^+, \mu_j^- \}_{j=1}^n$. The only poles that appear are kinematic poles, when $\mu_j^+\rightarrow \mu_k^-$, and they can be easily regularized (see Section~\ref{regularization}). Therefore we do not need to evaluate the form factors precisely at the $\{\mu_j^+, \mu_j^- \}_{j=1}^n$ that follows from the solutions of the Bethe equations. In fact we can take now $\{\mu_j^+, \mu_j^- \}_{j=1}^n$ to be independent free parameters (macroscopic excitations) that we denote $\{p_j, h_j\}_{j=1}^n$. For each choice of $\{p_j, h_j\}_{j=1}^n$ there is a number $\exp\left(\sum_{j=1}^n \delta S[\vartheta; p_j, h_j]\right)$ (with $\delta S$ defined in~\eqref{S_diff}) of microscopic states which share the same form factor up to finite size corrections. In order then to use macroscopic variables we need to multiply the form factors at fixed $\{p_j, h_j\}_{j=1}^n$ times the number of microscopic states that are characterized by the same macroscopic excitations. This allows to define the thermodynamic limit of the form factors for smooth filling functions $\vartheta$
\begin{equation}
|\langle \vartheta | \hat{\rho} | \vartheta, \{ h_j \to p_j\}_{j=1}^n \rangle |= \lim_{\text{th}} \Big( L^{n}  |\mathcal{F}(\vartheta; \{\mu^-\}, \{\mu^+\})|\Big) \times \exp\left(\sum_{j=1}^n \delta S[\vartheta; p_j, h_j]\right)  .\label{FF_TL}
\end{equation}
Moreover we can recast the sum over the macroscopic rapidites of the excitations into integrals by taking a special care of the divergences encountered (as is done in section~\ref{regularization})
\begin{equation}
\sum_{\mu_1^+< \ldots < \mu_n^+}  \sum_{\mu_1^-< \ldots < \mu_n^-}=  \frac{1}{n!^2}\sum_{\{\mu_j^+,\mu_j^-\}_{j=1}^n} =  L^{2n} \frac{1}{n!^2} \left( \int_{-\infty}^{\infty} dp_j  \rho_h(p_j)   \fint_{-\infty}^{\infty} d h_j \rho(h_j) \right).
\end{equation}
where $ \fint$ denotes the finite part of the integral, defined for a generic function $f(h)$ with a pole in $h=p$ as
\begin{equation}
  \fint_{-\infty}^{\infty} d h f(h) = \lim_{\epsilon \to 0^+}  \int_{-\infty}^\infty dh f(h + i \epsilon) - \pi i \underset{h=p}{\rm res} f(h).
\end{equation}

With this notation we can then write the correlation functions as a sum over all the possible excitations on the thermodynamic state~\cite{1990_Korepin_NPB_340}
\begin{align}
  & \langle\hat{\rho}(x,t)  \hat{\rho}(0,0) \rangle = \sum_{n=0}^{\infty} \frac{1}{n!^2} \left( \int_{-\infty}^{\infty} d p_j \rho_h(p_j)\:   \fint_{-\infty}^{\infty} d h_j \rho(h_j) \right)|\langle \vartheta | \hat{\rho} | \vartheta, \{ h_j \to p_j \}_{j=1}^n \rangle |^2 \nn \\&
\times  \prod_{j=1}^n  \exp \Big( {ix   (k(p_j) -k(h_j) ) -it (\omega(p_j) - \omega(h_j))}  \Big) , \label{corr_func_TL}
\end{align}
where the momentum and energy of the excitation follows \eqref{exc_momentum} and \eqref{exc_energy} respectively.

Eq.~\eqref{corr_func_TL} reminds the LeClair-Mussardo formula that appears in the context of integrable field theories~\cite{1999_LeClair_NPB_552}
\begin{align}
  & \langle\hat{\rho}(x,t)  \hat{\rho}(0,0) \rangle = \sum_{n=0}^{\infty} \frac{1}{n!^2} \left( \int_{-\infty}^{\infty} d p_j \rho_h(p_j)    \fint_{-\infty}^{\infty} d h_j \rho(h_j) \right)|\langle 0 | \hat{\rho} | 0, \{ h_j \to p_j \}_{j=1}^n \rangle_{FT} |^2 \nn \\&
\times  \prod_{j=1}^n  \exp \Big( {ix   (k(p_j) -k(h_j) ) -it (\omega(p_j) - \omega(h_j))} \Big)  \label{corr_func_LM}.
\end{align}
The main difference between eqs.~\eqref{corr_func_TL} and~\eqref{corr_func_LM} are the form factors used. The field theoretic form factors simply come from excitations over a structure-less vacuum.  Here, eq.~\eqref{corr_func_TL} suggests that the concept of vacuum is not appropriate for the strongly correlated systems. The form factors still depend \emph{explicitly} on the properties of the averaging state (through the filling function $\vartheta$). Conceptually this difference is responsible for the insufficiency of the field theoretical approach to the two-point correlation function (contrary to the one-point functions where the field theory approach is correct)~\cite{2002_Saleur_NPB_602, 2002_Alvaredo_NPB_636}. On the computational level this was explicitly shown in~\cite{1742-5468-2010-11-P11012}.

\subsection{Finite size corrections}

In order to compute the thermodynamic limit of the form factors we need to characterize the density of the particles and the back-flow function up to order $1/L$. This is due to existence of products in eq.~\eqref{ff} which are of order $N$ and in the thermodynamic limit can yield finite contributions from order $1/L$ terms. The derivation is similar to the one presented in~\cite{2012_Shashi_PRB_85} for the ground state distribution. Therefore here we simply state the result highlighting few differences between the two cases and for the details we refer to the section IV.B of \cite{2012_Shashi_PRB_85}.

The filling function $\vartheta(\lambda)$, as well as the particle density $\rho(\lambda)$, comes from the thermodynamic limit of a certain class of Bethe states. Let $\{\lambda\}_{j=1}^N$ be the Bethe roots of one of these states. Let us consider Bethe equations \eqref{bethe} and define a variable $x$ that satisfies
\begin{align}
  \lambda(x) = 2\pi x - \frac{1}{L}\sum_{k=1}^N \theta(\lambda(x) - \lambda_k).
\end{align}
Clearly $\lambda(I_j/L) = \lambda_j$ but we allow here $x$ to take any real value. Therefore $dx/d\lambda$ has a meaning of a number of possible quantum numbers in the range $d\lambda$. Thus
\begin{align}
  \frac{dx}{d\lambda} = \rho_t(\lambda).
\end{align}
and from the Euler-Maclaurin formula we have
\begin{align}
  \rho_t(\lambda) = \frac{1}{2\pi} + \frac{1}{2\pi}\int_{\lambda_1}^{\lambda_N} K(\lambda-\mu) \rho(\mu) d\mu + \frac{1}{2L}\left(K(\lambda-\lambda_N) - K(\lambda - \lambda_1)\right),
\end{align}
where $\lambda_{1,N}$ are the smallest and largest rapidities respectively. Incorporating the $1/L$ in the boundaries of the integral yields
\begin{align}
   \rho_t(\lambda) = \frac{1}{2\pi} + \frac{1}{2\pi}\int_{-q}^{q} K(\lambda-\mu) \rho(\mu) d\mu,
\end{align}
with $q_L = \lambda_1 + 1/(2L\rho_p(\lambda_1))$ and $q_R = \lambda_N + 1/(2L\rho_p(\lambda_N))$. In the thermodynamic limit we have $q_{R,L}\rightarrow\pm\infty$ and it is convenient to separate the thermodynamic part from the finite size corrections. Before doing so, let us note that we can bound the finite-size corrections from above by choosing $q = \textrm{min}(|q_L|, q_R)$. We have
\begin{align}
 \rho_t(\lambda) =& \frac{1}{2\pi} + \frac{1}{2\pi}\int_{-\infty}^{\infty} K(\lambda-\mu) \rho(\mu) d\mu - \frac{1}{2\pi}\int_{q}^{\infty} \nn \left(K(\lambda-\mu)+K(\lambda+\mu)\right) \rho(\mu) d\mu. 
\\ =& \frac{1}{2\pi} + \frac{1}{2\pi}\int_{-\infty}^{\infty} K(\lambda-\mu) \rho(\mu) d\mu\nn\\ &- \frac{1}{2\pi}\int_{0}^{\infty} \left(K(\lambda-\mu-q)+K(\lambda+\mu+q)\right) \rho(\mu+q) d\mu.
\end{align}
The last term that controls the finite-size corrections can be easily bounded (where $M$ is a positive constant)
\begin{align}
  \frac{1}{2\pi}\int_{q}^{\infty} \left(K(\lambda-\mu)+K(\lambda+\mu)\right) \rho(\mu) d\mu \leq M\frac{\rho(q)}{2\pi},
\end{align}
and the finite size corrections are proportional to $\rho(q)$. For the energy of the state to be finite we require $\rho(\lambda) \sim \lambda^{-3-\epsilon}$ for large $\lambda$ (c.f. \eqref{energy_cont}). The boundary $q$ itself is a monotonically increasing function of $N$. For the particle density to spread over the whole real line, rather than to accumulate in the final interval of it, we should have $q \sim N^{1+\delta}$ with $\delta > 0$. Therefore $\rho(q) \sim N^{-3-\gamma}$ with $\gamma>0$ and the finite-size corrections are at least of order $1/L^3$ and thus are negligible in the further analysis.
We have
\begin{align}
  \rho_t(\lambda) &= \frac{1}{2\pi} + \frac{1}{2\pi}\int_{-\infty}^{\infty} K(\lambda-\mu) \rho(\mu) d\mu + \mathcal{O}(1/L^3).
\end{align}

The finite-size corrections to the particle and hole rapidities follow from analogous computations as presented in \cite{2012_Shashi_PRB_85} and are given by
\begin{align}\label{finite_size_excitations}
  \mu_0^+ &= 2\pi \frac{I^+}{L} - \int_{-\infty}^{\infty} \theta(\mu_0^+ - \lambda) \rho(\lambda) d\lambda, \;\;\;\;\;\;
  \mu_{1/L}^+ = -\frac{F(\mu_0^+)}{L\rho_t(\mu_0^+)}, \\
  \lambda_0^- &= 2\pi \frac{I^-}{L} - \int_{-\infty}^{\infty} \theta(\lambda_0^- - \lambda) \rho(\lambda) d\lambda, \;\;\;\;\;\;
  \lambda_{1/L}^- = 0.
\end{align}
For the auxiliary rapidity $\mu^-$ we have
\begin{align}
  \mu_0^- = \lambda_0^-, \;\;\;\;\;\;
  \mu_{1/L}^- = -\frac{F(\lambda_0^-)}{L\rho_t(\mu_0^-)},
\end{align}
We can consider now finite-size corrections to the back-flow function. The thermodynamic limit is given by eq. \eqref{backflow}. The leading finite size corrections are
\begin{align}
  2\pi F_{1/L}(\lambda) &= \left(\int_{q_R}^{\infty} + \int_{-\infty}^{q_L}\right)d\mu\, K(\lambda,\mu) \vartheta(\mu) F(\mu) \nonumber\\
  &- \frac{1}{L}\left[ K(\lambda,\mu_0^+)\left( \frac{F(\lambda)}{\rho_t(\lambda)}-\frac{F(\mu_0^+)}{\rho_t(\mu_0^+)}\right) - K(\lambda,\mu_0^-)\left( \frac{F(\lambda)}{\rho_t(\lambda)}-\frac{F(\mu_0^-)}{\rho_t(\mu_0^-)}\right)\right] \nn\\
 & + \frac{1}{2L} \int_{-\infty}^{\infty} d\mu\, \rho(\mu)K'(\lambda-\mu) \left(\frac{F(\lambda)}{\rho_t(\lambda)} - \frac{F(\mu)}{\rho_t(\mu)}\right)^2.
\end{align}
The first integral can be estimated in the following way
\begin{align}
\left|\int_{q_R}^{\infty} d\mu\,K(\lambda,\mu) \vartheta(\mu) F(\mu)\right| &\leq \int_{q_R}^{\infty} d\mu \left|\,K(\lambda,\mu) \vartheta(\mu) F(\mu)\right| \nonumber\\ &\leq |\vartheta(q_R)| \int_{q_R}^{\infty} d\mu \left|K(\lambda,\mu) F(\mu)\right|.
\end{align}
But $\vartheta(q_R)$ is proportional to $\rho(q_R)$ which is of order $1/L^3$. The same holds for the other integral between $-\infty$ and $q_L$. Therefore both integrals can be neglected in the leading order and
\begin{align}
  2\pi L F_{1/L}(\lambda) =& - \left[ K(\lambda,\mu_0^+)\left( \frac{F(\lambda)}{\rho_t(\lambda)}-\frac{F(\mu_0^+)}{\rho_t(\mu_0^+)}\right) - K(\lambda,\mu_0^-)\left( \frac{F(\lambda)}{\rho_t(\lambda)}-\frac{F(\mu_0^-)}{\rho_t(\mu_0^-)}\right)\right] \nonumber\\
&+ \frac{1}{2} \int_{-\infty}^{\infty} d\mu\, \rho(\mu)K'(\lambda-\mu) \left(\frac{F(\lambda)}{\rho_t(\lambda)} - \frac{F(\mu)}{\rho_t(\mu)}\right)^2.
\end{align}
Finally the density of the excited state is
\begin{align}
  \rho_{t,ex}(\mu_j) = \rho_t(\lambda_j) + \frac{1}{L}\left(F'(\lambda_j) - F(\lambda_j)\frac{\rho_t'(\lambda_j)}{\rho_t(\lambda_j)} \right).
\end{align}
This completes the list of formulas required to take properly the thermodynamic limit of the form factors. This is achieved in the next section.

\section{Thermodynamic limit for smooth distribution of rapidities} \label{thermodynamiclimit}

In this section we calculate the thermodynamic limit of the finite size (normalized) form factors~\eqref{micro_ff} (see also~\eqref{ff} and~\eqref{norm} for explicit formulas). We are interested in the leading term in $1/L$ so with the equivalence $\sim$ we denote that we are neglecting extra sub-leading correction in $1/L$.  We proceed as in~\cite{2012_Shashi_PRB_85} since the two calculations share many common steps. Introducing the short-hand notation
\begin{equation}
\mu^{}_{ij} = \mu^{}_i - \mu^{}_j ,
\end{equation}
we start from an intermediate expression for the thermodynamic limit of the form factors given by (c.f. eq. 88 in \cite{2012_Shashi_PRB_85})
\begin{align}\label{starting}
|\mathcal{F}_N| \sim& \prod_{i=1}^n\left( \frac{F_L(\lambda^-_i)}{L(\rho_L(\lambda^-_i)\rho_{ex}(\mu^+_i))^{1/2}(\lambda_i^- - \mu_i^+)}\right)\frac{i \Delta k \: {\det_N}(\delta_{jk} + U_{jk})}{(V_p^+ - V_p^-){\rm Det}\left(1-\frac{\hat{K \vartheta}}{2\pi} \right)}\nn\\
&\times \left[\prod_{j,k} \left(\frac{(\lambda_j - \mu_k +ic)(\lambda_j - \mu_k + ic)}{(\lambda_{jk} + ic)(\mu_{jk} + ic)}\right)^{1/2}\right] \nn\\ &
\times \left\{\prod_j\frac{{\rm sin}(\pi F_L(\lambda_j))}{\pi F_L(\lambda_j)} \prod_{j\neq k}\left(\frac{\lambda_{jk} \mu_{jk}}{(\mu_j - \lambda_k)^2}\right)^{1/2}\right\}\nn\\ \times
&{\rm exp}\left[\int_{-\infty}^\infty d\lambda \vartheta(\lambda) \left(\frac{\pi F_L(\lambda) {\rm cos}(\pi F_L(\lambda))}{ {\rm sin}(\pi F_L(\lambda))} - 1\right) \right] \nn \\ & \times \exp \left[ \int_{-\infty}^{\infty} d\lambda \vartheta(\lambda) \left[ \left( F_L'(\lambda) - \frac{F_L(\lambda)\rho_t'(\lambda)}{2\rho_t(\lambda)}\right) + \frac{1}{2}F_L'(\lambda) \right]\right] .
\end{align}
where ${\rm Det}( 1- \frac{K \vartheta }{2 \pi})$ is the Fredholm determinant of the kernel
\begin{align}
 -\left[\frac{K \vartheta }{2 \pi}\right](\mu,\nu) = -\frac{1}{2 \pi}\frac{2c}{(\mu - \nu)^2 + c^2} \vartheta(\nu) ,
\end{align}
In eq. \eqref{starting} there are three groups of elements which are still written for a finite system. We denote them as
\begin{align}
  M_1 &= \prod_{j,k} \left(\frac{(\lambda_j - \mu_k +ic)(\lambda_j - \mu_k + ic)}{(\lambda_{jk} + ic)(\mu_{jk} + ic)}\right)^{1/2}, \\
  M_2 &= \prod_j\frac{{\rm sin}(\pi F_L(\lambda_j))}{\pi F_L(\lambda_j)} \prod_{j\neq k}\left(\frac{\lambda_{jk} \mu_{jk}}{(\mu_j - \lambda_k)^2}\right)^{1/2}, \\
  \Theta &= \frac{\det_N (\delta_{jk} + U_{ij})}{V_p^+ - V_p^- }.
\end{align}
The thermodynamic limit of them requires some work. Calculation of $M_1$ is exactly the same as for the ground state form factors and thus we do not reproduce it here. For the details we refer again to \cite{2012_Shashi_PRB_85}. On the other hand the term $M_2$ has a manifestly different thermodynamic limit and is responsible for different size dependence of the ground state (critical) form factors and the finite entropy state form factors. Computations are presented in the next section. The thermodynamic limit of $\Theta$ was computed in \cite{2012_Shashi_PRB_85} only for a specific type of excitations. As we require here the form factors for a generic particle-hole excitations we have to generalize the previous calculations. This is done in the subsequent section.

\subsection{\texorpdfstring{Evalutation of $M_2$}{Evaluation of M(2)}}
We focus here on the evaluation of the double products given by
\begin{equation}
M_2 = \prod_j \frac{\sin(\pi F_L(\lambda_j))}{\pi F_L(\lambda_j)} \prod_{j \neq k =1}^N \left(\frac{\lambda_{jk} \mu_{kj}}{(\mu_k - \lambda_j)^2} \right)^{1/2} ,
\end{equation}
which present formal differences in the thermodynamic limit when the states is described by a smooth distribution or when the distribution is discontinuous as for the ground state. Differently from the ground state situation this term is not expected to produce power law divergences in the system size as $1/L^\alpha$.

\begin{figure}
\centering
\includegraphics[scale=0.3]{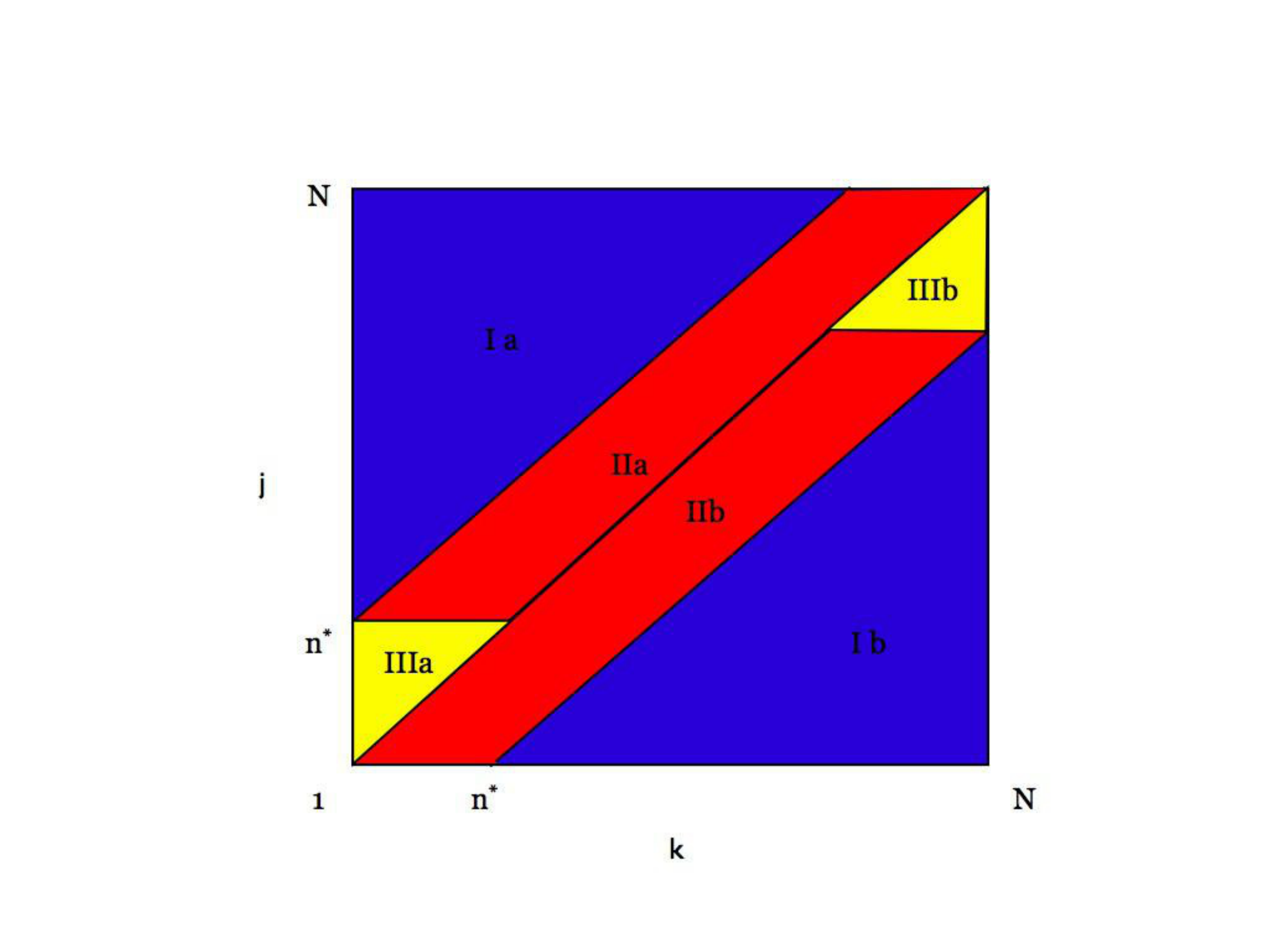}\label{M2_regions}
\caption{Schematic of range of $\lambda$'s. Cutoff parameter n* is used to separate regions I, II. Figure is taken from \cite{2012_Shashi_PRB_85} and there are two extra regions (IIIa and IIIb) depicted. These are the regions in which the two rapidities are close two each other and simultaneously they are close to the edge of the distribution. For the smooth distribution of rapidities there is no edge and consequently the region III does not appear in the present calculations. Ultimately the existence or lack of region III determines the scaling of the form factors with the system size $L$. It is the region III that leads to the fractional power of the system size for the ground state form factors and consequently to the criticality of zero temperature 1D Bose gas. This shows an intimate connection between the distribution of rapidities and the form of the correlation function. (Figure reproduced from~\cite{2012_Shashi_PRB_85}, courtesy of A. Shashi).}
\end{figure}

Following \cite{2012_Shashi_PRB_85} we decompose the product in three pieces
\begin{equation}
M= T'' \times T_{holes} \times T_{particles} ,
\end{equation}
depending on which rapidities we let the sum run over.
\begin{align}
T'' &=\prod_{j \neq k}'' \left(\frac{\lambda_{jk} \mu_{kj}}{(\mu_k - \lambda_j)^2} \right)^{1/2} \nonumber \\
&\sim \prod_{j \neq k}'' \left(1 + \frac{F_L(\lambda_k)}{L \rho_t(\lambda_k)(\lambda_j - \lambda_k)} \right)^{-1/2} \left(1 - \frac{F_L(\lambda_k)}{L \rho_t(\lambda_k)(\lambda_j - \lambda_k)} \right)^{-1/2} \nonumber\\
&\times \left(1 + \frac{F_L(\lambda_k)}{L \rho_t(\lambda_k)(\lambda_j - \lambda_k)}  -\frac{F_L(\lambda_k)}{L \rho_t(\lambda_j)(\lambda_j - \lambda_k)} \right)^{-1/2} ,
\end{align}
where by $\prod_{j \neq k}''$ we denoted the product where we excluded the particles excitations $\{ \mu^+_j \}_{j=1}^n$ but we included the holes $\{ \mu^-_j \}_{j=1}^n$. When the two rapidities get closer, i.e.\
when $j \in [k-n^*, k+ n^*]$ where $n$ is a given sub-extensive cut-off such that $ n \propto L^{1- \alpha}$ with $\alpha<1/2$, then we substitute for the difference between the two
\begin{equation}\label{Expanded_rap}
L \rho_t(\lambda_j )(\lambda_j- \lambda_k)= \frac{j-k}{\vartheta(\lambda_j)}  +  \frac{ (j-k)^2 \partial_\lambda \rho_t(\lambda_j)}{2 L \rho(\lambda_j)^2} ,
\end{equation}
while for all the other $j$ we can just exchange the sum for an integral over the rapidities (see figure~\ref{M2_regions}).
The cut-off $n^*$ delimits the region where the approximation \eqref{Expanded_rap} start to break down, which corresponds to a distance in rapidities
\begin{equation}
|\lambda_j - \lambda_k | \equiv \nu^*(\lambda) = \frac{n^*}{L \rho(\lambda)} + \mathcal{O}(n^*/L^2) .
\end{equation}
We denote the two regions in $\lambda_j - \lambda_k$ separated by the cut-off as the region $I$ (smooth part) and $II$ (discrete part) where the approximation \eqref{Expanded_rap} is valid (IIa where $j < k$ and IIb where $k< j$ ). $T''$ is then given by the product of these three terms
\begin{equation}
T'' = T_I \times T_{IIa} \times T_{IIb} ,
\end{equation}
For the fist term we have the following
\begin{align}
  \log T_1 =& \left[\frac{1}{2} \sum_{j \neq [k - n^* , k+n^*]} \frac{F_L(\lambda_j) F_L(\lambda_k)}{L^2 \rho_t(\lambda_j)\rho_t(\lambda_k)(\lambda_j- \lambda_j)} \right] \nn\\
  \sim& \frac{1}{2} \left( \int d\lambda \vartheta(\lambda) \int_{-\infty}^{\lambda - \nu^*(\lambda)} \!\!d \mu \vartheta(\mu) \frac{F(\lambda) F(\mu)}{(\lambda - \mu)^2} +\int d\lambda \vartheta(\lambda) \int_{\lambda + \nu^*(\lambda)}^{\infty} \!\!d\mu \vartheta(\mu) \frac{F(\lambda) F(\mu)}{(\lambda - \mu)^2} \right) \nn \\
  =& -\frac{1}{4} \int d\lambda  \int d \mu  \frac{(F_L(\lambda) \vartheta(\lambda)- F_L(\mu) \vartheta(\mu))^2}{(\lambda - \mu)^2}  \nn \\
  &+ \frac{1}{2} \int d\lambda \vartheta(\lambda)^2 F_L(\lambda)^2\left( \int_{-\infty}^{\lambda - \nu^*(\lambda)} d \mu \frac{1}{(\lambda - \mu)^2} + \int_{\lambda + \nu^*(\lambda)}^{\infty} d \mu  \frac{1}{(\lambda - \mu)^2} \right) \nn \\
  =& -\frac{1}{4} \int d\lambda  \int d \mu  \frac{(F_L(\lambda) \vartheta(\lambda)- F_L(\mu) \vartheta(\mu))^2}{(\lambda - \mu)^2}
  + \int d\lambda \vartheta(\lambda)^2 F_L(\lambda)^2 \frac{1}{\nu^*(\lambda)}  \nn \\
  =& -\frac{1}{4} \int d\lambda  \int d \mu  \frac{(F_L(\lambda) \vartheta(\lambda)- F_L(\mu) \vartheta(\mu))^2}{(\lambda - \mu)^2} \nn \\
  &+ \frac{L}{n^*}\int d\lambda \vartheta(\lambda)^2 F_L(\lambda)^2 \rho(\lambda)  + \mathcal{O}\left(\frac{L}{(n^*)^2} \right).
\end{align}
The computation in the sector II can be done analogously as  in \cite{2012_Shashi_PRB_85}  leading to
\begin{align}
\log T_{II}
= & -  \int d\lambda  \frac{\rho'_t(\lambda)}{\rho_t(\lambda)}\frac{ F(\lambda)} {2\rho(\lambda) \partial_\lambda(F(\lambda)\vtheta(\lambda))}  \frac{\partial }{\partial \lambda} \log \left( \frac{\pi F(\lambda) \vtheta(\lambda)}{\sin \pi F(\lambda) \vtheta(\lambda)} \right) \\& - \frac{L}{ n^*}\int d\lambda \vartheta(\lambda)^2 F_L(\lambda)^2 \rho(\lambda)  + \mathcal{O}(L/(n^*)^2) ,
\end{align}
where the cut-off depended part cancels exactly the one in $T_I$ leading to a cut-off independent result.

Now we are left only with the computation of $T_{particle}$ and $T_{hole}$ defined as
\begin{equation}
T_{hole} \sim \prod_{j}' \prod_{k=1}^n \left( 1 - \frac{F_L(\lambda_j)}{L \rho(\lambda_j)(\lambda_j - \lambda_k^-) } \right)^{-1} ,
\end{equation}
\begin{equation}
T_{particle} \sim \prod_{j}' \prod_{k=1}^n \left( 1 - \frac{F_L(\lambda_j)}{L \rho(\lambda_j)(\lambda_j - \lambda_k^-) } \right) ,
\end{equation}
where the product $\prod_{j}'$ runs respectively over all the rapidities in the state but not over the holes(particle).
Proceeding as in \cite{2012_Shashi_PRB_85} and using the fact that for a smooth distribution particle and holes are always arbitrarily close to an extensive number of rapidities of the state, we obtain
\begin{equation}
T_{hole} =  \prod_{k=1}^n \frac{\sin \pi F(\mu_k^-) \vtheta(\mu_k^-)}{\pi F(\mu_k^-)  \vtheta(\mu_k^-) } \exp\left( \int_{-\infty}^{\infty} d\lambda \vartheta(\lambda) \: \frac{F(\lambda) - F(\mu_k^-) }{(\lambda - \mu_k^-)  } \right) .
\end{equation}
For $T_{particle}$ we have the same result
\begin{equation}
T_{particle} = \prod_{k=1}^n  \frac{\pi F(\mu_k^+)   \vtheta(\mu_k^+) }{  \sin \pi F(\mu_k^+)  \vtheta(\mu_k^+)} \exp\left( - \int_{-\infty}^{\infty} d\lambda \vartheta(\lambda) \: \frac{F(\lambda) - F(\mu_k^+) }{(\lambda - \mu_k^+)  } \right) .
\end{equation}

Finally we can then write for the whole factor $M_2$ in the thermodynamic limit
\begin{align}
  M_2 & = \exp\left[ -  \int d\lambda  \frac{\rho'_t(\lambda)}{\rho_t(\lambda)}\frac{ F(\lambda)} {2\rho(\lambda) \partial_\lambda(F(\lambda)\vtheta(\lambda))}  \frac{\partial }{\partial \lambda} \log\left( \frac{\pi F(\lambda) \vtheta(\lambda)}{\sin \pi F(\lambda) \vtheta(\lambda)} \right) \right] \nn \\&
  \times \prod_{k=1}^n \frac{\sin \pi F(\mu_k^-) \vtheta(\mu_k^-)}{\pi F(\mu_k^-)  \vtheta(\mu_k^-) }  \frac{\pi F(\mu_k^+)   \vtheta(\mu_k^+) }{  \sin \pi F(\mu_k^+)  \vtheta(\mu_k^+)}  \nn \\& \times
  \exp\left[ \int_{-\infty}^{\infty} d\lambda \vartheta(\lambda) \: \frac{F(\lambda) - F(\mu_k^-) }{(\lambda - \mu_k^-)  } )  - \int_{-\infty}^{\infty} d\lambda \vartheta(\lambda) \: \frac{F(\lambda) - F(\mu_k^+) }{(\lambda - \mu_k^+)  } \right] \nn \\& \times
  \exp\left[ \frac{-1}{4} \int d\lambda  \int d \mu  \frac{(F_L(\lambda) \vartheta(\lambda)- F_L(\mu) \vartheta(\mu))^2}{(\lambda - \mu)^2} \right] ,\label{M2_final}
\end{align}

\subsection{Fredholm determinant}
We are left with the problem of computing the thermodynamic limit of the determinant
\begin{equation}\label{theta}
\Theta = \frac{\det_N (\delta_{jk} + U_{jk})}{V_p^+ - V_p^- }
\end{equation}
where the matrix $U$ is given in \eqref{matrix_U} and $\lambda_p$ is an arbitrary number. Analogously as is done in \cite{2012_Shashi_PRB_85} we can take the limit $\lambda_p \to \infty$ leading to
\begin{equation}
\Theta = \frac{i c}{2 \Delta k } \det \left(\delta_{jk} + \frac{i (\mu_j - \lambda_j)}{V_j^+ - V_j^- } \prod_{m \neq j} \frac{\mu_m - \lambda_j}{\lambda_m - \lambda_j} \left( \frac{2 c}{(\lambda_j - \lambda_k)^2 + c^2} - \frac{2}{c} \right) \right)
\end{equation}
It is useful to consider a vector
\begin{equation}
a_j = \frac{i (\mu_j - \lambda_j)}{V_j^+ - V_j^-} \prod_{m \neq j} \frac{\mu_m - \lambda_j}{\lambda_m - \lambda_j}.
\end{equation}
With this notation the determinant in \eqref{theta} is expressed as
\begin{align}
  \det_N \left(  \delta_{ij}+ A_{ij}\right),
\end{align}
with the matrix $A$ given by
\begin{align}
  A_{jk} = a_j \left(K(\lambda_j-\lambda_k) - \frac{2}{c} \right)
\end{align}
Depending on $j$ the vector $a_j$ has a different scaling behaviors with the system size
 \begin{align}
   a_j \sim \begin{cases}
     \mathcal{O}(1/L), &\lambda_j \notin \{\lambda_j^-\}_{j=1}^n\\
     \mathcal{O}(1), &\lambda_j \in \{\lambda_j^-\}_{j=1}^n
   \end{cases}.
\end{align}
We can use this property to simplify the computation of the determinant.
We denote by $\tilde{a}_j$ a vector $a_j$ in which we substitute $\mu_j$ by $\mu_j^-$. Correspondingly we define a matrix $\tilde{A}_{jk}$ with $\tilde{a}_j$ as a prefactor instead of $a_j$. Thus the matrix elements of $\tilde{A}_{jk}$ are all of $\mathcal{O}(1/L)$. By $B_{jk}$ we denote the difference of the two matrices: $B_{jk} = A_{jk} - \tilde{A}_{jk}$. Note that the matrix $B_{jk}$ has only $n$ non-zero rows corresponding to the excited rapidites $\{ \lambda_j^-\}_{j=1}^n$. Using standard proprieties of the determinant and assuming the matrix $\delta_{ij} +\tilde{A}_{ij} $ is invertible we can recast the determinant in a product of the determinant of an $N\times N$ matrix and the determinant of an $n\times n$ one
\begin{equation}\label{two_det}
  \det_N (\delta_{ij}+ \tilde{A}_{ij}+ B_{ij} ) =  \det_N( \delta_{ij} +\tilde{A}_{ij} ) \times \det_n \left(   \delta_{ij} + \sum_{k=1}^N B_{ik} \left(1+ \tilde{A}\right)^{-1}_{kj}\right),
\end{equation}
where the indices $i,j$ in the second determinant run only over the $n$ excited rapidities
and $1$ denotes here the identity matrix. We can now take the thermodynamic limit. The first determinant becomes a Fredholm determinant ${\rm Det}(1 + \hat{A})$ with the kernel
\begin{equation} \label{kernelA}
  \hat{A}(\lambda, \mu) = \tilde{a}(\lambda)\left( K(\lambda-\mu) - \frac{2}{c}\right),
\end{equation}
where $\tilde{a}(\lambda_j)$ is the thermodynamic limit of $\rho(\lambda_j){a}_j$. Standard computations give \cite{2012_Shashi_PRB_85}
\begin{align}
  \tilde{a}(\lambda ) = &  \frac{\vtheta(\lambda)F(\lambda)}{\Gamma[1 + \vtheta(\lambda)F(\lambda)]  \Gamma[1 - \vtheta(\lambda)F(\lambda)]} \underset{\mu^-_k \neq \lambda}{\prod_{k=1}^n} \frac{ \mu_k^+ - \lambda}{\mu_k^- - \lambda}\exp \left[ - {\rm PV}\! \int d\mu  \frac{\vtheta(\mu) F(\mu)}{\mu - \lambda} \right]
    \nn \\& \times
\left[2 {\rm Im} \left( \prod_{k=1}^n \frac{\mu_k^+ - \lambda + i c}{\mu_k^- - \lambda+ ic}   \exp\left[- \int d\mu \frac{\vtheta(\mu)F(\mu)}{\mu - \lambda + ic} \right] \right)\right]^{-1}. \label{a_tilde_derivation}
\end{align}
where the product $\underset{\mu^-_k \neq \lambda}{\prod_{k=1}^n} $ runs over all the holes (particles) except the $j-$th one when $\lambda = \mu_j^{-}$ for any $j=1,\ldots, n$. With ${\rm PV} \int$ we denoted the principal value of the integral
\begin{equation}
{\rm PV} \int dx \frac{f(x)}{x} = \lim_{\epsilon \to 0} \left( \int_{-\infty}^{-\epsilon}dx \frac{f(x)}{x} + \int_{\epsilon}^{\infty} dx \frac{f(x)}{x} \right)
\end{equation}
 The second line of eq.~\eqref{a_tilde_derivation} can be still simplified. We separate the exponential term into real and imaginary parts and for the imaginary part use the integral equation for the backflow \eqref{backflow} to obtain
\begin{align}
  &\exp \left(-\int d\mu \frac{\vartheta(\mu) F(\mu)}{\mu - \lambda +ic}\right) = \exp\left(-\frac{1}{2c} \int d\mu (\mu-\lambda)\vartheta(\mu) F(\mu) K(\lambda - \mu)\right) \nonumber\\ &\times \exp\left(i\pi F(\lambda) -\frac{i}{2}\sum_{k=1}^n \left(\theta(\lambda-\mu_k^+) - \theta(\lambda - \mu_k^-) \right) \right).
\end{align}
From the definition of $\theta(\lambda) = 2{\rm atan}(\lambda/c)$~\eqref{phase_shift} and an identity \
\begin{align}
  \exp\left(2i{\rm atan}(x)\right) = \frac{1 + ix}{1 - ix},
\end{align}
it follows that
\begin{align}
  &{\rm Im}\left[ \prod_k^n\frac{\mu_k^+ - \lambda +ic}{\mu_k^- - \lambda + ic}\exp\left(-\int d\mu \frac{\vartheta(\mu) F(\mu)}{\mu - \lambda +ic}\right) \right] = \sin \pi F(\lambda) \prod_{k=1}^n\left(\frac{K(\mu_k^- - \lambda)}{K(\mu_k^+ -\lambda)} \right)^{1/2}
\nonumber\\
&\times
\exp\left(-\frac{1}{2c} \int d\mu (\mu-\lambda)\vartheta(\mu) F(\mu) K(\lambda - \mu)\right) .
\end{align}
Using Euler's reflection formula for $\Gamma$ functions
\begin{align}
  \Gamma(1-z)\Gamma(1+z) = \frac{\pi z}{\sin \pi z},
\end{align}
we obtain
\begin{align}
  &\tilde{a}(\lambda) = \frac{\sin[ \pi \vartheta(\lambda) F(\lambda)]}{2\pi \sin [ \pi F(\lambda)]} \nonumber\\& \times  \frac{\prod_{k=1}^n \mu_k^+ - \lambda}{\underset{\mu^-_k \neq \lambda}{\prod_{k=1}^n}\mu_k^- - \lambda} \prod_{k=1}^n\left(\frac{K(\mu_k^+ -\lambda)}{K(\mu_k^- - \lambda)} \right)^{1/2}
   \exp \left[ \frac{c}{2}{\rm PV}\! \int d\mu  \frac{\vtheta(\mu)F(\mu) K(\mu - \lambda)}{\mu - \lambda} \right] .
\end{align}

In the $n \times n$ determinant in \eqref{two_det} we note that we can neglect the $1/L$ corrections to $B$ since $\lim_{\text{th}} \frac{n}{L}=0$. We introduce the matrix $W = A (1+\tilde{A})^{-1}$ as the solution of the following equation
\begin{equation}
W_{ij} + \sum_{k=1}^N W_{ik} \tilde{A}_{kj}= A_{ij},\;\;\;i,j=1,\dots,n,
\end{equation}
which in the thermodynamic limit becomes a linear integral equations for the function $W(\lambda , \mu)$
\begin{equation}\label{kernelW}
  W(\lambda, \mu) + \int_{-\infty}^\infty d\alpha W(\lambda, \alpha) \tilde{a}(\alpha) \left(K(\alpha - \mu) - \frac{2}{c} \right)    =    b( \lambda)\left(K(\lambda - \mu) - \frac{2}{c} \right)   ,
\end{equation}
with the vector $b(\lambda)$ given by
\begin{align}
  b(\lambda) & =- \frac{\sin [\pi \vartheta(\lambda) F(\lambda)]}{2\pi \vartheta(\lambda) F(\lambda) \sin [\pi F(\lambda)]} \nonumber\\
  &\times \frac{\prod_{k=1}^n \mu_k^+ - \lambda}{\underset{\mu^-_k \neq \lambda}{\prod_{k=1}^n}\mu_k^- - \lambda} \prod_{k=1}^n\left(\frac{K(\mu_k^+ -\lambda)}{K(\mu_k^- - \lambda)} \right)^{1/2}
\exp \left[ \frac{c}{2}{ \rm PV} \int d\mu\frac{\vartheta(\mu) F(\mu) K(\lambda - \mu)}{\lambda - \mu}\right],
\end{align}

Putting everything together we have then
\begin{equation}
  \limth \det_N (\delta_{ij}+ \tilde{A}_{ij}+ B_{ij} ) =  {\rm Det}( 1 +\hat{A} ) \det_n \left(  \delta_{ij} + W(\mu_i^-,\mu_j^-) \right),
\end{equation}
and the determinant part of the form factors is then expressed as
\begin{equation}
  \Theta = \frac{i c}{2 \Delta k} {\rm Det}( 1+ \hat{A})   {\rm det}_n \left(  \delta_{ij} + W(\mu_i^-,\mu_j^-) \right). \label{theta_final}
\end{equation}
Note that the Fredholm determinant is still a function of the excitations and not only of the the averaging state. This unfortunately poses still serious problems to the computation of correlation functions. However from the numerical point of view its evaluation can be effectively approximated with the very first terms of its expansion in powers of the trace.

\subsection{Final result}
We report here the final expression for the thermodynamic limit of the form factors of the density operator between the representative state given by a smooth distribution $\rho(\lambda)$ and a number $n$ of particle-hole excitations. Most of the remaining computations can be carried out exactly as is done in \cite{2012_Shashi_PRB_85} by simply rescalling the shift function as $\tilde{F}(\lambda) = \vtheta (\lambda)F(\lambda)$.
Combining the partial thermodynamic limit of the form factors from eq. \eqref{starting} with the results for $M_1$, $M_2$ (\eqref{M2_final} and $\Theta$ (\eqref{theta_final} we find the form factors between a thermodynamic state $|\vartheta\rangle$ and one of its excited states with $n$ particle-holes, as defined in \eqref{FF_TL}, to be
\begin{align}\label{FFd_final_expression}
  & |\langle \vartheta | \hat{\rho} | \vartheta, \{ h_j \to p_j\}_{j=1}^n \rangle |= \nn \\&
  \frac{c}{2} \left[\prod_{k=1}^n \frac{F(h_k)}{ (\rho_t(p_k) \rho_t(h_k))^{1/2} } \frac{\pi \tilde{F}(p_k)    }{  \sin \pi \tilde{F}(p_k)  } \: \frac{\sin \pi \tilde{F}(h_k)}{\pi \tilde{F}(h_k)   } \right] \nn \\&  \times
  \prod_{i,j=1}^n \left[\frac{(p_i - h_j + i c)^2}{(h_{i,j} + ic)(p_{i,j} + ic)} \right]^{1/2} \frac{\prod_{i< j =1}^n  h_{ij} p_{ij}}{\prod_{i , j} (p_i - h_j)}
  \det_n \left(  \delta_{ij} + W(h_i,h_j) \right) \nn \\& \times \exp\left(- \frac{1}{4} \int d\lambda \int d\mu \left( \frac{\tilde{F}(\lambda) - \tilde{F}(\mu)}{\lambda - \mu}\right)^2   - \frac{1}{2} \int d\mu d \lambda \left( \frac{\tilde{F}(\lambda)\tilde{F}(\mu)}{(\lambda - \mu + i c)^2}\right) \right) \nn \\& \times
  \exp\left( \sum_{k=1}^n  {\rm PV} \int_{-\infty}^{\infty} d\lambda  \: \frac{\tilde{F}(\lambda)  (h_k - p_k)  }{(\lambda - h_k) ( \lambda - p_k)   }+  \int d\lambda \frac{\tilde{F}(\lambda) (p_k - h_k)}{(\lambda - h_k + i c) (\lambda - p_k + i c)}  \right)
  \nn \\& \times \frac{{\rm Det}\left(1 + \hat{A} \right)}{{\rm Det}\left(1  - \frac{K \vartheta}{2 \pi }\right)} \exp\left(\sum_{j=1}^n \delta S[\vartheta; p_j, h_j]\right),
\end{align}
with the kernels $\hat{A}$ and $W$ given respectively in \eqref{kernelA} and \eqref{kernelW}. The form factors are now completely characterized by thermodynamic data. Knowing the $\vartheta(\lambda)$ function we can find the density $\rho_t(\lambda)$. Specifying the rapidities of the excitations $\{h_j\rightarrow p_j\}_{j=1}^n$ the back-flow function $F(\lambda|\{h_j\rightarrow p_j\}_{j=1}^n)$ and the form factor itself follows. Note that in order to have a complete resolution of identity we need to include also the diagonal form factor with $n=0$
\begin{equation} \label{diagonal}
|\langle \vartheta | \hat{\rho} | \vartheta \rangle |= D
\end{equation}
where the density of particles can be chosen to be unitary $D=1$.

The expression \eqref{FFd_final_expression} is complicated and the meaning of many terms is rather obscure. The main difficulty is hidden in the Fredholm determinant which depends on the excitations and a factorization of it is still not possible. In order to have some insight on the structure of the form factors it is interesting to consider the small density limit. That is we let $\vartheta(\lambda) \approx 0$ and obtain
\begin{align}
  & |\langle 0 | \hat{\rho} | 0, \{ h_j \to p_j\}_{j=1}^n \rangle |=
\frac{c}{2} \left[\prod_{k=1}^n  \sum_{l=1}^n (\theta(p_l - h_k)  - \theta(h_l - h_k)  )\right] \nn \\&  \times
\prod_{i,j=1}^n \left[\frac{(p_i - h_j + i c)^2}{(h_{i,j} + ic)(p_{i,j} + ic)} \right]^{1/2} \frac{\prod_{i< j =1}^n  h_{ij} p_{ij}}{\prod_{i , j} (p_i - h_j)}
 \det_n \left(  \delta_{ij} + W(h_i,h_j) \right)\label{FFd_zero_density},
\end{align}
where we used that $\rho_t(\lambda) = 1/(2\pi) + \mathcal{O}(\vartheta(\lambda))$ and
\begin{align}
  F(\lambda) = \frac{1}{2\pi} \sum_{k=1}^n \left(\theta(p_k - \lambda) - \theta(h_k - \lambda)\right) + \mathcal{O}(\vartheta(\lambda)).
\end{align}
The matrix $W(h,p)$ \eqref{kernelW} also simplifies. The kernel $\hat{A}$ becomes small and we obtain an explicit expression for $W(\mu, \lambda)$
\begin{align}
  W(\lambda, \mu) =  b(\lambda)\left(K(\lambda - \mu) - \frac{2}{c} \right) + \mathcal{O}(\vartheta(\lambda))
\end{align}
with
\begin{align}
  b(\lambda) = \frac{-1}{2\sin\left[\frac{1}{2} \sum_{k=1}^n\left(\theta(p_k -\lambda)-\theta(h_k - \lambda) \right) \right]} \prod_{k=1}^n \frac{K^{1/2}(h_k-\lambda)}{K^{1/2}(p_k-\lambda)}  \frac{\prod_{k=1}^n (p_k - \lambda)}{\underset{h_k \neq \lambda}{\prod_{k=1}^n} (h_k - \lambda)}.
\end{align}
In the case of 1 particle-hole excitation the form factor simplifies to
\begin{equation} \label{1ph_small_density}
 |\langle 0 | \hat{\rho} | 0, h \to p \rangle |  = \frac{1}{2}\frac{\theta(p-h)}{ (p-h)} \left((p-h)^2 + c^2\right)^{1/2} ,
\end{equation}
Note that the form factor \eqref{1ph_small_density} describes a process of creating a particle-hole excitation in a low density state. Therefore is very different from the form factor \eqref{ff} for $N=1$. The later equals
\begin{equation}
 |\langle \mu| \hat{\rho} |\lambda\rangle| = c,
\end{equation}
and describes the process of exciting a single particle state with momentum $\lambda$ to momentum $\mu$ (Since these are single particle states the momentum is equal to the rapidity.). This shows that particle-hole excitations over the averaging state cannot be identified with particle creation over the vacuum in the field theory. Note that contrary to the relativistic field theory \cite{MussardoBOOK} there is no crossing symmetry that would allow to transform the hole in the ket state into a particle in the bra state in eq.~\eqref{1ph_small_density}.

\section{Regularization of the divergences} \label{regularization}
To compute correlation functions we need to perform an integration over all possible values of the rapidites of the excitations.
The form factors \eqref{FFd_final_expression} have however a singularity whenever $h_j = p_k$ and they are finite only when we consider only one single particle-hole  $n=1$ with $p \to h$, when the form factor becomes indeed diagonal. Therefore we need to be careful while rewriting the sums as integrals. The aim of this section is to show how this can be done. Let us start with the finite size form of the correlation function where we already neglect sub-leading corrections \eqref{corr_func_TL}
\begin{align}
& \langle \hat{\rho}(x,t) \hat{\rho}(0) \rangle = \sum_{n=0}^\infty \frac{1}{n!^2} \prod_{j=1}^n \left[   \frac{1}{L} \sum_{p_j} \frac{1}{L}\sum_{h_j} \right] \\&
\times    |\langle \vartheta | \hat{\rho} | \vartheta, \{ h_j \to p_j\}_{j=1}^n \rangle |^2 e^{\sum_{j=1}^n  \left[ - i x  (k(p_j) - k(h_j)) - i t ( \omega(p_j)  - \omega(h_j))\right]}.
\end{align}
The sum over particle and holes rapidites transforms into a product of integrals under a proper regularization. The idea, already introduced to regularize the field theory form factors in \cite{1742-5468-2010-11-P11012}, is to write the sum over the holes as a complex integral over all the values that the holes rapidites can take for a finite (but large) $L$ using \eqref{finite_size_excitations}
\begin{equation}
L Q(h) =L \left( h + \int d\lambda \theta(h - \lambda) \rho(\lambda) \right) = 2 \pi I_j,
\end{equation}
where $\{ I_j \}$ are all the quantum numbers of the averaging state at some large fixed system size $L$.
With a help of $Q(h)$  we can write the sum of a function $f(z)$ over all the values of hole rapidity $h$ as
\begin{align}
 \frac{1}{L}\sum_{h} f(h) =&   \sum_{I_j}  \oint_{I_j} \frac{dz}{2 \pi} \frac{f(z) Q'(z)}{ e^{i L Q(z)} -1} \nonumber\\
=&  \left(\int_{\mathbb{R} - i \epsilon} -\int_{\mathbb{R} + i \epsilon} \right)\frac{f(z) Q'(z)}{ e^{i L Q(z)} -1} \frac{dz}{2 \pi} -   \sum_{\text{poles(f)} \in \Gamma_\epsilon} \oint dz \frac{f(z) Q'(z)}{ e^{i L Q(z)}-1} -  \sum_{r_j \not \in \{ I_j \}  } f(z_j),
\end{align}
where the first integrals are taken on a single contour including the poles in $Q(z) = 2 \pi I_j$ where $I_j$ are all the possible quantum numbers of the hole. In the second step we modified the sum over all these contours in the integral over the line above and below the real axes. In order to do that we need to subtract extra poles that we do not want to include. One type of them are the poles of $f(z)$ in the stripe $\Gamma_\epsilon$ delimited by the two imaginary lines. Other poles are located at the values $z$ such that $Q(z)= 2 \pi r_j$ with $r_j$ not a quantum number of the averaging state (where holes cannot be created). When $L \to \infty$ only the integral above the real line survives the limit (since $Q(z)$ is monotonic in $z$) leading to
\begin{equation}
\frac{1}{L}\sum_{h} f(h)   =  \int_{\mathbb{R} + i \epsilon} {f(z) \rho(z)}{} dz -  \pi i  \sum_{\text{res(f)} \in \Gamma_\epsilon} {f(z) \rho(z)}{ }.
\end{equation}
 If now we impose that $f(z)$ has only a double pole in $z=p$ we can then rewrite the sum in terms of the finite part of the integral over $h$
 \begin{equation}
\frac{1}{L}\sum_{h} f(h)  = \lim_{\epsilon \to 0^+}  \int_{-\infty}^\infty dh f(h + i \epsilon) - \pi i \underset{h=p}{\rm res} f(h) = \: \fint_{-\infty}^{\infty} d h f(h).
\end{equation}
In order to compute the finite part is then useful to compute the limit  $p_j \to h_j$ of the form factors \eqref{FFd_final_expression}
\begin{align} \label{FF_recursion}
  \frac{|\langle \vartheta | \hat{\rho} | \vartheta, \{ h_j \to p_j\}_{j=1}^n \rangle |}{|\langle \vartheta | \hat{\rho} | \vartheta, \{ h_j \to p_j\}_{j=1}^{n-1} \rangle |} {=} \frac{F(h_n)}{\rho_t(h_n)(p_n - h_n)}  + \mathcal{O}(p_n - h_n).
\end{align}
where the back-flow is now computed as the sum of the other back-flows for the residual excitation \eqref{back-sum-flow}
\begin{equation}
  F\left(\lambda\,|\, \{(\mu_j^+, \mu_j^-)\}_{j=1}^n\right) = \sum_{j=1}^{n-1} F\left(\lambda\,|\, \mu_j^+, \mu_j^-\right) .
\end{equation}

\section{\texorpdfstring{Dynamical structure factor in $1/c$ expansion}{Dynamical structure factor in 1/c expansion}} \label{expansion}
We consider here the expansion in $1/c$ of the dynamical structure factor, defined as the Fourier transform of  the density-density correlation
\begin{align}\label{dsf}
&S(q, \omega) =\int dx dt \: e^{i q x - i \omega t} \langle \rho(\lambda) |  \hat{\rho}(x,t)  \hat{\rho}(0,0) | \rho(\lambda) \rangle
\nn \\&
=(2 \pi)^2 \sum_{n=0}^\infty \frac{1}{n!^2}\left[ \prod_{j=1}^n \int_{-\infty}^{\infty} d p_j \rho_h(p_j)   \fint_{-\infty}^{\infty} d h_j \rho(h_j)    \right] \delta\left(q-\sum_{j=1}^n (k(p_j) - k(h_j))\right)  \nonumber\\
& \times \delta\left(\omega - \sum_{j=1}^n (\omega(p_j) - \omega(h_j) \right)   |\langle \vartheta | \hat{\rho} | \vartheta, \{ h_j \to p_j\}_{j=1}^n \rangle |^2  , \nonumber \\
\end{align}
for a generic thermal state at temperature $T=\beta^{-1}$ and density $D=1$.
Expanding at the first order in $1/c$ the only relevant form factors are the ones with only 1 particle-hole excitation $p,h$
\begin{align}
 & |\langle \vartheta | \hat{\rho} | \vartheta, h \to p \rangle | \nn \\&
 =
  \frac{1}{2 \pi } \frac{1 + \frac{2}{ c} }{(\rho_t(h) \rho_t(p))^{1/2}}
 \left[1 - \frac{(p - h)^2}{\pi c} {\rm PV} \int d\lambda \frac{\vartheta(\lambda)}{(\lambda - p)(\lambda - h)} \right] + \mathcal{O}(1/c^2),
\end{align}
since the ones with two or more particle-hole excitations contribute at the order $1/c^2$ or higher. The filling fraction for a thermal state at temperature $\beta$, including the $1/c$ correction, is given by
\begin{align}
 \vartheta(\lambda)  = \frac{1 + \frac{2}{c}}{1 + e^{\beta(\lambda^2 - h)}} ,
\end{align}
with $h$ the chemical potential fixing the density $D=1$ of the gas.
In the 1 particle-hole spectrum dynamical structure factor at $S(q,\omega)$ is given in terms of a single form factor with energy $\omega$ and momentum $q$ times the density of states, which is simply the Jacobian of the transformation from the rapidities of the excitations to the energy and momentum variable
\begin{align}
p^2   - h^2   & = \omega,   \\
p  - h  & = q \Big( 1 + \frac{2}{c} \Big)^{-1},
\end{align}
which gives a Jacobian factor $\Big|\det \frac{\partial ( \omega, q)}{\partial( p, h ) }  \Big|= 2 q (1+2/c)^{-1}$ with the rapidities of the excitations given by
\begin{align}
& p =  \frac{q}{2 (1 + 2/c)} + \frac{\omega (1 + 2/c)}{2 q} , \\
& h = - \frac{q}{2 (1 + 2/c)}+  \frac{\omega (1 + 2/c)}{2 q}.
\end{align}

We obtain then an expression for the thermal dynamical structure factor up to $1/c^2$ corrections
\begin{align}\label{dynamical_final}
S(q, \omega) &=  (2\pi)^2 \frac{1 + \frac{2}{c}}{2 q }   \Big[\rho_h( p) \rho( h)|\langle \vartheta | \hat{\rho} | \vartheta, h \to p \rangle |\Big]\nonumber\\
&=\frac{2 }{\pi} \left( \pi \frac{1 + \frac{6}{c}}{4 q} +  \frac{1}{2 c} {\rm PV}\int \frac{\vartheta(\lambda + p)  - \vartheta(\lambda + h)}{\lambda  }  \right) {\vartheta(h)\left( 1 - \vartheta(p)\right)} + \mathcal{O}(1/c^2).
\end{align}
Using
\begin{align}
  1-e^{-\beta\omega}=\frac{\vartheta(h) - \vartheta(p)}{\vartheta(h)(1- \vartheta(p))}.
\end{align}
we obtain the same result as in \cite{2005_Brand_PRA_72} (where here we have chosen unitary density $D=1$). Note that the limit $T \to 0$ can be easily recovered from \eqref{dynamical_final}.  The same is believed to be true for all the orders in $1/c$ of the correlation functions. This is a non-trivial statement since the procedure to obtain the form factors when the averaging state is the ground state and when is a thermal state are manifestly different.

The example here is carried on for a thermal state, however this result can be extended to any filling fraction $\vartheta(\lambda)$ including for example the saddle point state after a quench in the Lieb-Liniger model \cite{2014_DeNardis_PRA_89}

\section{\texorpdfstring{Numerical evaluation of the dynamical structure factor}{Numerical evaluation}} \label{numerics}

\begin{figure}
\centering
\includegraphics[scale=0.52]{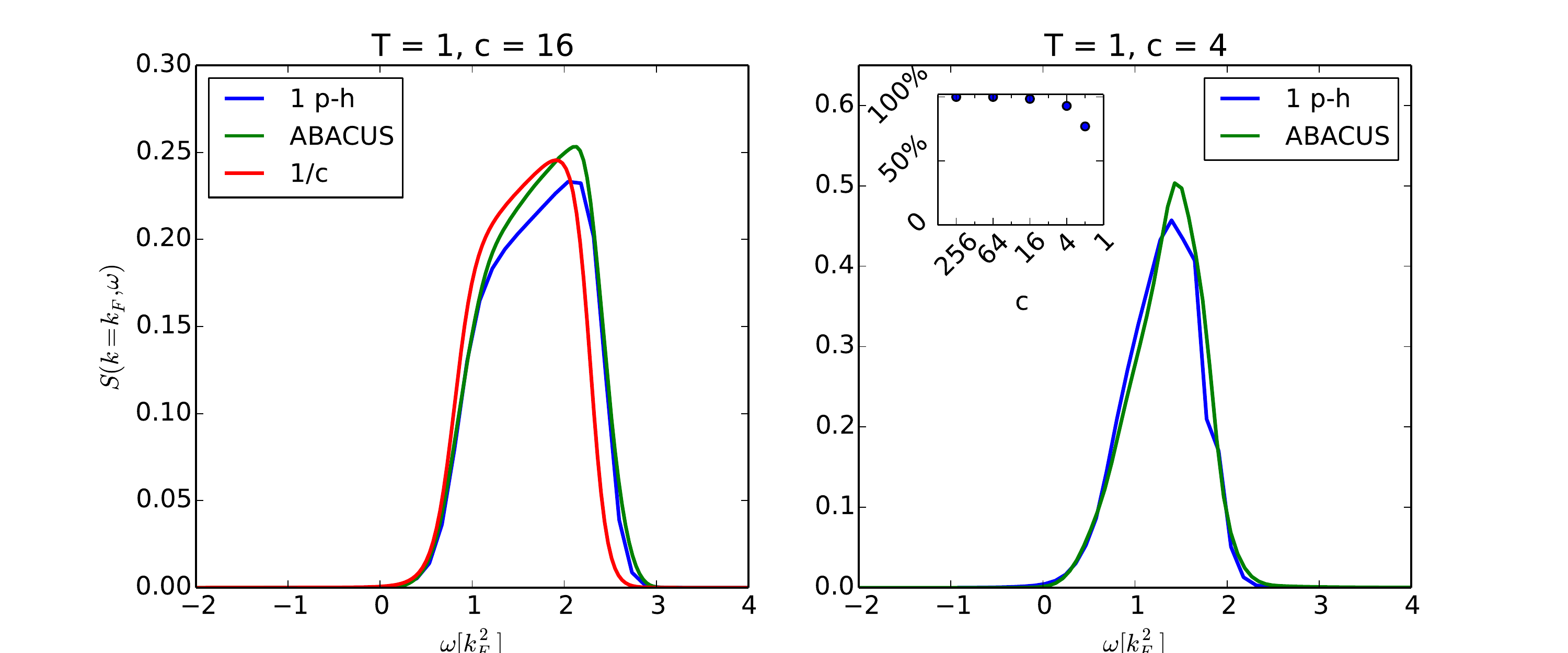}
\caption{Correlation function $S(k,\omega)$ at $T=1$, unitary density $D=1$ and with $c=16$ (\emph{on the left}) and with $c=4$ (\emph{on the right}). We plot it as a function of energy $\omega$ for fixed value of momentum $k=k_F$. The 1 particle-hole approximation $S_{1ph}(q, \omega)$ (blue) fits well the ABACUS results (green) and misses only on the correlation weight which is expected to come from multiple particle-hole excitations. For $c=16$ we plot also the results of the $1/c$ expansion (red). Inset in the right figure shows the percentage of the f-sum rule \eqref{fsumrule} saturation for some values of the interaction parameter $c$.}
\label{fig2}
\end{figure}

\begin{figure}
\centering
\includegraphics[scale=0.52]{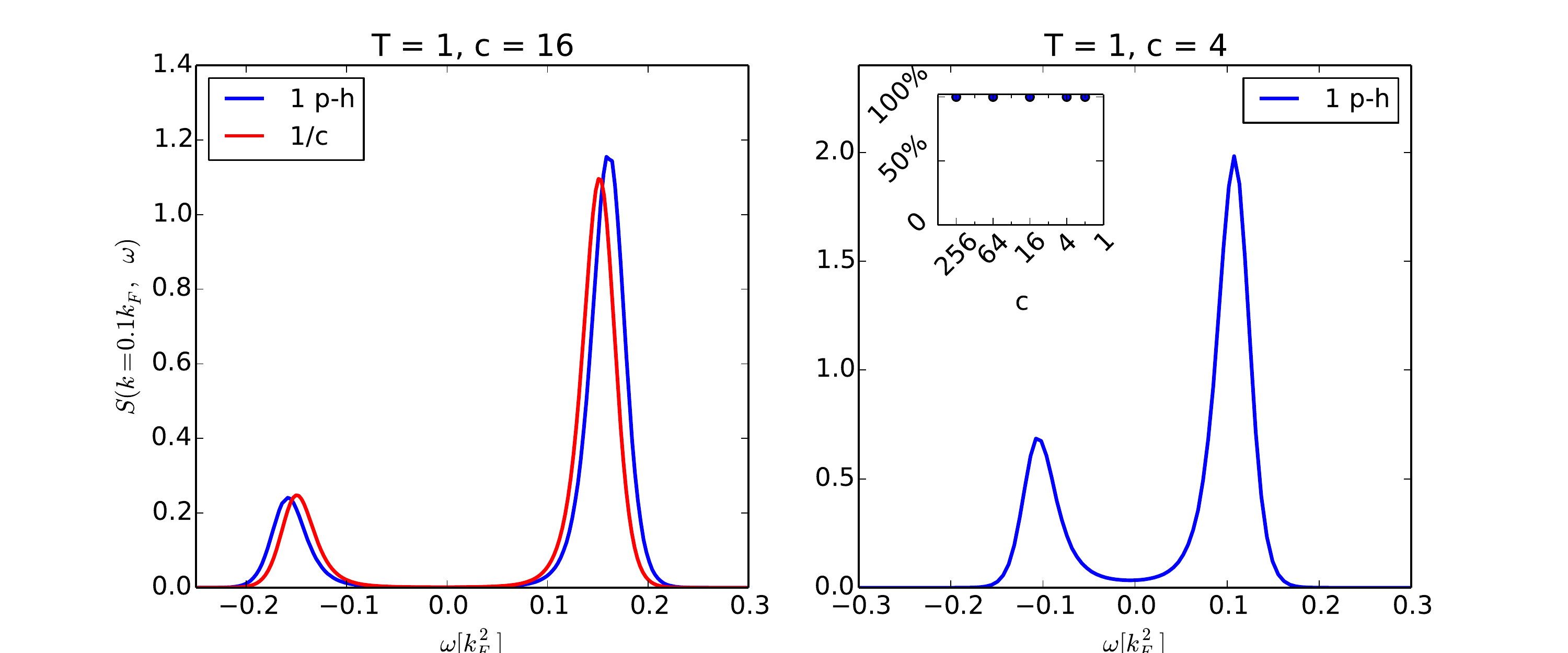}
\caption{Correlation function $S(k,\omega)$ at $T=1$, unitary density $D=1$ and with $c=16$ (\emph{on the left}) and with $c=4$ (\emph{on the right}) as a function of energy $\omega$ for fixed value of momentum $k=k_F/10$. For  $c=16$ we plot also the result of the $1/c$ expansion. The f-sum rule (as plotted inside the inset) is saturated exactly (up to a numerical precision of $\sim 1\%$) since the 1 particle-hole approximation becomes exact at small momenta. Lower value of momentum allows for development of a second peak, at negative energies, due to the detailed balance relation $S(k,-\omega) = S(k, \omega) \exp(-\omega/T)$~\cite{LL_StatPhys2_BOOK}. Note that at higher momenta the correlation function is shifted towards higher energies and the negative energy peak becomes practically invisible (See figure~\ref{fig2}).}
\label{fig3}
\end{figure}

The dynamical structure factor \eqref{dsf} can be computed through numerical evaluations of the exact formula \eqref{FFd_final_expression}. The sum over all the possible number of excitations $n = 1, 2 , \ldots$ requires a great numerical effort, mainly due to the complicated structure of the form factors. To simplify the problem we focus here only on the simplest excitations ($n=1$) consisting of a single particle-hole pair. This leads to an approximate expression for the correlation function which is shown in figures~\ref{fig2} and~\ref{fig3}. As in the $1/c$ section, for concreteness we limit ourselves to thermal equilibrium correlations.

The approximation of the correlation function to a single particle-hole pair becomes exact in the limit of large interactions (c.f. previous section) and in the limit of very small momentum $k$.\footnote{This is the same idea as presented in study of the dynamic structure factor of the XXZ spin chain at small momentum~\cite{2007_Pereira_JSTAT_8}.} As the interaction is decreased and momentum is increased we expect the approximation to become worse. To quantify how far the resulting correlation function is from the true one we compare our results with an exact numerical evaluation of the correlation function in a finite system~\cite{PhysRevA.89.033605} via the ABACUS algorithm~\cite{2009_Caux_JMP_50}. This shows that even for values of $c\sim 1$, which go well beyond the $1/c$ expansion and at finite momentum $k=k_F$, the 1 particle-hole contribution captures the essential features of the dynamic structure factor. For $c\approx 4$ the kinetic and potential energy \eqref{H} of the ground state of the system are equal \cite{JCS_Comment} and thus the correlation function is the most difficult to compute.

Additionally we consider the f-sum rule~\cite{LL_StatPhys2_BOOK}, an exact equality obeyed by the dynamic structure factor for any fixed momentum $q$
\begin{equation} \label{fsumrule}
\int_{-\infty}^\infty \frac{d\omega}{2 \pi} \omega S(q,\omega) = D q^2.
\end{equation}
In the limit $c\to \infty$ or $k\rightarrow 0$ the 1 particle-hole spectrum is the full excitation spectrum for the density operator and consequently the f-sum rule is completely saturated by including only these types of excitations in the sum \eqref{dsf}. However as $c$ decreases with $k$ finite we observe that the f-sum rules is saturated only up to a certain precision and more excitations have to be taken into account in order to obtain the full correlation function. Again, even at values of $c \sim 1$  and $k\sim k_F$ the contribution of the 1 particle-hole excitations remains very significant (see insets of figures~\ref{fig2} and~\ref{fig3}).

The results of this section confirm that the form factors~\eqref{FFd_final_expression} can be directly used to compute the dynamic structure factor or in general the density-density correlation on a generic state with non-zero entropy. Moreover it asserts that the expansion in particle-hole excitation numbers is an effective method to compute the correlation function.

\section{Conclusions} \label{discussion}

In this work we studied the thermodynamic limit of the particle-hole form factors for the density operator of the 1D Bose gas. The computations presented here can be generalized to different operators (like the bosonic field operator $\Psi$) but also to other Bethe Ansatz solvable models for which the microscopic matrix elements are known such as the XXZ spin chain. These problems will be addressed in the future.
These form factors constitute the building blocks to compute thermal or post-quench equilibrium correlation functions in the thermodynamic limit at fixed density of particles. They also provide a first step towards the post-quench time evolution as recently done in \cite{me} for the Tonks-Girardeau ($c=\infty$) regime.

The final formula \eqref{FFd_final_expression} is valid in the thermodynamic limit and it is considerably simpler than its finite size version but still it is not suitable to obtain close-form expressions of correlation functions. The Fredholm determinant of the kernel $\hat{A}$ is a non-trivial functions of the excitations parameters $\{ p_j, h_j\}_{j=1}^n$ and we were not able to obtain further simplifications. A fully factorized expression of the form factors involving a simple almost factorized part depending only on the excitation parameters is still under research.

We computed the exact thermodynamic dynamical structure factor including only the 1 particle-hole excitations over a thermal state. This approximation is qualitatively different from the usual perturbative one or the low energy limit. For example the perturbation theory in $1/c$ breaks at $c\sim 10$ yielding unphysical, negative values of the correlation~\cite{2005_Brand_PRA_72}, while the (non-linear) Luttinger liquid theory is not able to reproduce the exact shape of the correlation function~\cite{2012_Imambekov_RMP_84}. We showed that a thermodynamic Bethe Ansatz approach with only single particle-hole excitations produces a good estimate of the density correlations of the system for a wide range of values of the interaction parameter and momentum. Therefore the effect of the extra particle-hole excitations is mainly to increase  the weight of the correlation at large momentum.

Another interesting point is to compare our result with similar ones for the thermodynamic limit of one-point functions of the Lieb-Liniger model obtained from the non-relativistic limit of the sinh-Gordon model \cite{2009_Kormos_PRA_81}. As shown in \cite{1742-5468-2011-11-P11017} the large volume limit of the diagonal form factor obtained by Bethe Ansatz \eqref{diagonal} can be expressed as a LeClair-Mussardo series \cite{1999_LeClair_NPB_552} of the elementary form factors obtained via the bootstrap program \cite{MussardoBOOK,2009_Kormos_PRA_81}. How to extend this relation to two-point functions remains to be clarified. The Bethe Ansatz approach, presented in this work, might shed a light on this important problem of the Quantum Integrable Field Theories.

Finally, following \cite{2011_Shashi_PRB_84} where a relation between the form factors and the prefactors of the Lutinger liquid correlation functions at zero temperature was established, it would be interesting to see whether such simple relations also exist at finite temperature or even out-of-equilibrium. The result of \cite{2011_Kozlowski_JSTAT_P03019} where low temperature correlation functions were studied seem to suggest that such relations might exists. This will be also a subject of a further research.

\ack
We are very grateful to J.-S. Caux for his support and critical comments and to R. Konik for a stimulating and encouraging discussion.
We acknowledge useful and inspiring discussions with S. Eli\"{e}ns, G. Mussardo and H. Saleur.
\noindent J. De Nardis acknowledges support from the Netherlands Organisation for Scientific Research (NWO). M. Panfil acknowledges support from the Foundation for Fundamental Research on Matter (FOM) at the early stage of this work.

\noindent This work was supported by ERC under the Starting Grant n. 279391 EDEQS.

\section*{References}

\bibliographystyle{iopart-num}
\bibliography{smoothening}

\end{document}